# Maximum likelihood reconstruction for the Daya Bay Experiment

**Dongmei Xia (On behalf of the Daya Bay Collaboration)**

Institute of High Energy Physics, Beijing, 100049, China

xiadm@ihep.ac.cn

**Abstract**. The Daya Bay Reactor Neutrino experiment is designed to precisely determine the neutrino mixing angle $\theta_{13}$. In this paper, we report a maximum likelihood (ML) method to reconstruct the vertex and energy of events in the anti-neutrino detector, based on a simplified optical model that describes light propagation. We calibrate the key parameters of the optical model with $^{60}$Co source, by comparing the predicted charges of the PMTs with the observed charges. With the optimized parameters, the resolution of the vertex reconstruction is about 25cm for $^{60}$Co γ's.

## 1. Introduction to the optical model.

The antineutrino detector (AD) of the Daya Bay experiment has three nested cylindrical volumes separated by concentric acrylic vessels [1]. The innermost volume holds 20t of Gd-liquid sintilltor[2] as the antineutrino target. The middle volume is filled with 21t liquild scitillator which is the gamma catcher. There are 192 8-inch PMTs mounted on eight ladders installed along the circumference and within the mineral oil volume, which is the outer volume of AD. Two reflective panels with a film of Enhanced Specular Reflected (ESR) are placed at the top and bottom of the outer volume to increase the photon-statistics and improve the uniformity of the energy response. Three automated calibration units (ACU-A,ACU-B, ACU-C) are mounted at the top of AD. Each ACU contains a LED as well as two scaled capsules with the radioactive source that can be lowered individually into the Gd-LS along either the centreline or inner edge, or in the LS.

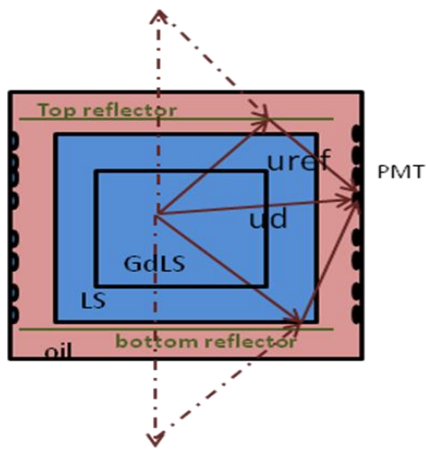

Fig. 1: An illustration of the optical model of the ML
reconstruction of AD for the Daya Bay experiment .

The optical model [3] based on the structure of the AD used in the ML reconstruction is illustrated in Fig.1. The predicted charge on each photomultiplier (PMT) is the sum of the photo-electrons (p.e) produced by the direct light ($\mu_d$) and the photo-electrons produced by the reflected light ($\mu_r$). The definition of $\mu_d$ and $\mu_r$ are in equation 1, where $\Phi$ is a normalization parameter; $f(\cos\theta_d)$ is the PMT angular response curve; $\theta_d$ is the angle between the PMT normal direction and the vector from PMT position pointing to source position; $R_d$ is the distance between PMT and radioactive source; $\lambda_a$ is the average attenuation length of the liquid scintillator and (LS) the Gd-LS ; $\varepsilon^{QE}$ is the PMT relative efficiency; $\mu_r$ is the sum of the charges produced by mirror sources, and the subscripts j, k denotes that light is the j-th order reflected by bottom reflector and the k-th order reflected by the top reflector; $\varepsilon_t, \varepsilon_b$ are the reflectivity for the top and bottom reflector, respectively; $R_{jk}$ is the distance from the mirror source to the PMT; $\theta_{jk}$ is the angle between the PMT normal direction and the vector from the PMT position pointing to the mirror source position. A sum of $\mu_d$ and $\mu_r$ gives the total expected charge.

$$\mu_d = \Phi \cdot \frac{f(\cos\theta_d)}{R_d^2} \cdot \exp\left(-R_d/\lambda_a\right) \cdot \varepsilon^{QE}$$
$$\mu_r = \Phi \cdot \sum_{j,k}\left(\varepsilon_b^j\right)\cdot\left(\varepsilon_t^k\right) \cdot \frac{f(\cos\theta_{j,k})}{R_{j,k}^2} \cdot \exp\left(-R_{j,k}/\lambda_a\right) \cdot \varepsilon^{QE} \quad (1)$$

Below are the key parameters that need be determined from calibration data:
- The average attenuation length $\lambda_a$.
- PMT angular response curve $f(\cos\theta_d)$.
- The top and bottom reflectivity $\varepsilon_t, \varepsilon_b$.
- The PMT relative efficiency $\varepsilon^{QE}$.

## 2. Calibrate the key parameters

The calibration of the key parameters is performed by requiring the expected charge distribution on the PMTs to agree with that observed in data. We use the 2.5MeV $\gamma$ emitted by $^{60}$Co for calibration, except for PMT relative efficiency calibration which use the low energy calibration source $^{68}$Ge .

*2.1. PMT relative efficiency*

The PMT relative efficiency is determined by counting the relative occupancy, when the calibration source is at the detector centre. A low energy calibration source $^{68}$Ge is chosen for calibrating. With this calibration source, at most one p.e is obtained by the PMT.

*2.2. Attenuation length、reflectivity、PMT angular response curve*

A $\chi^2$ function is built to calibrate the attenuation length $\lambda_a$、the reflectivity $\varepsilon_t, \varepsilon_b$ and the PMT angular response curve $f(\cos\theta_d)$ simultaneously, as shown in equation 2.

$$\chi^2 = \sum_{j}^{Num}\left(\sum_{i}^{192}\frac{(\overline{n_{ij}}-\mu_{ij})^2}{\overline{n_{ij}}}\right) \quad (2)$$

In equation 2, $\overline{n_{ij}}$ is the average observed charge, $\mu_{ij}$ is the expected charge and is a function about the parameters to be calibrated $\mu_{ij}(\lambda_a, \varepsilon_t, \varepsilon_b, f(\cos\theta))$, 192 is the number of PMTs, $Num$ is the total number of calibration sources that locate at different position in the AD, $f(\cos\theta_d)$ is parameterized as $f(\cos\theta) = p_0 + p_1\cos\theta + p_2\cos^2\theta$. Optimal parameters are determined by minimizing $\chi^2$

### 2.3. Performance of the parameters

Figure 2 shows the ratio of observed charge to expected charge as a function of the incident angle $\theta$ and the distance from the PMT to the radioactive source. $\theta$ is the angle between the PMT normal direction and the vector from PMT position pointing to source position. The performance of the optical model with the optimized parameters is much better than that with the initial parameters as figure 2 indicates. The initial parameters were measured independently before the assembling of AD.

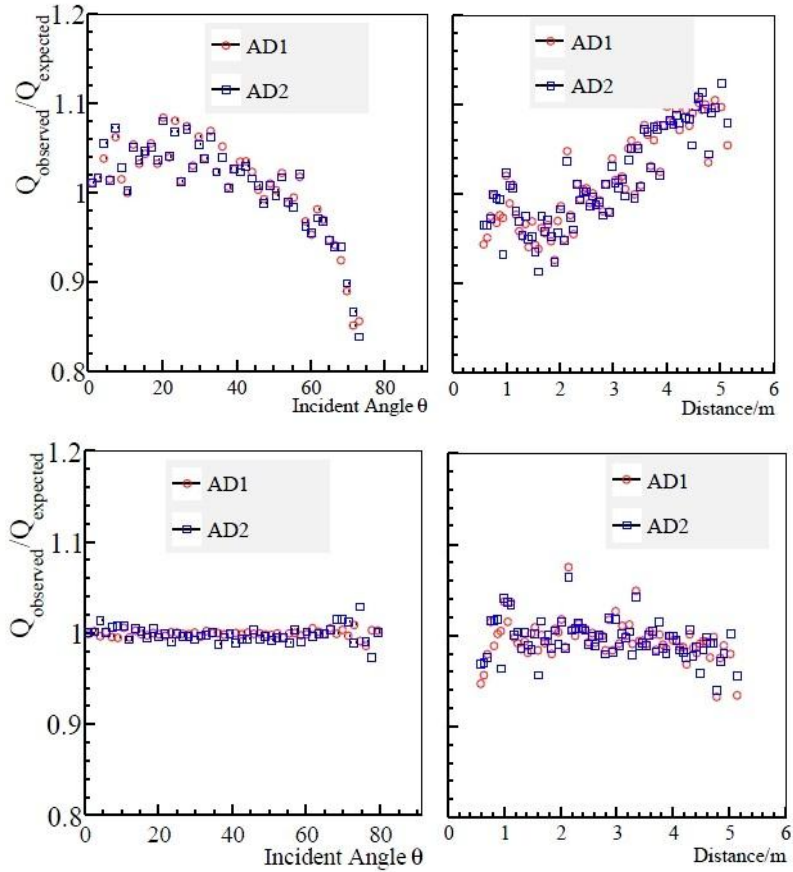

Fig 2: Top: The ratio of observed charge to the expected charge as a function of $\theta$ and distance for the old parameters; Bottom: The ratio of observed charge to expected charge as a function of $\theta$ and distance for the optimized parameters.

## 3. Performance of the Reconstruction

The performance of the ML reconstruction is studied by using the Am-C neutron source, the source was deployed in the detector along various vertical axes and radial directions.

### 3.1. Energy reconstruction

The accuracy of the energy reconstruction is investigated by comparing the peaks of the reconstructed energy and the true energy of neutron capture events. The energy peaks are determined by fitting the energy spectrum with Double Crystal Ball function as shown in figure 3.

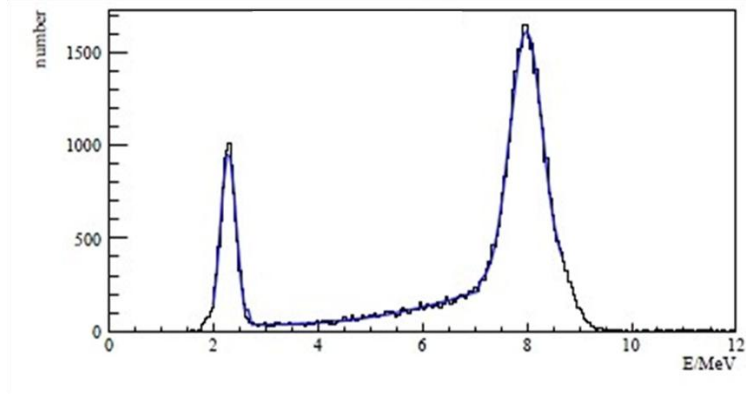

Fig.3.: Energy distribution of the Am-C neutron events.

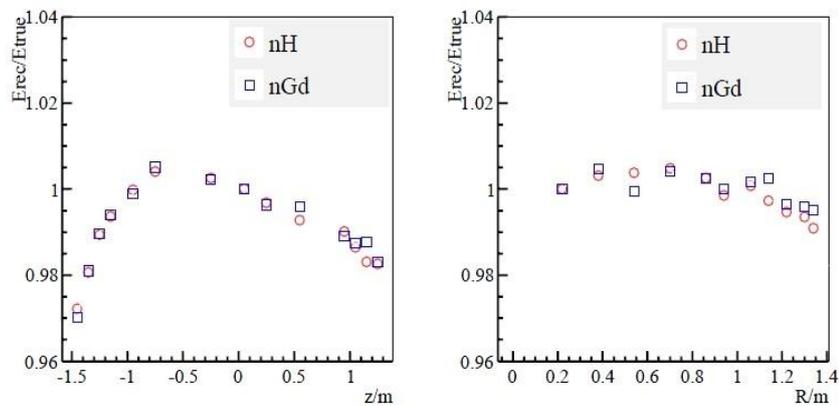

Fig.4.:The vertical (left) and radial (right) energy non-uniformity after ML reconstruction.

From figure 4, the non-uniformity along vertical direction of the energy reconstruction is within 4% for the neutron capture on Hydrogen (nH) events and the neutron capture on Gadolinium (nGd) events. The reconstructed energy is uniform along radial direction .

### 3.2. Vertex reconstruction

The bias of the vertex reconstruction is defined as the mean value of vertical and radial differences between the reconstructed vertex and the true vertex. The resolution of the vertex reconstruction is defined as the sigma of the vertex distribution. The bias increases when events are close to the top and bottom reflector. The maximum Z bias is about 20cm and the bias in X,Y is within 10cm along vertical direction. The bias is within 5cm along radial direction, as shown in figure5. The resolution of the ML reconstruction is within 25cm as shown in figure 6.

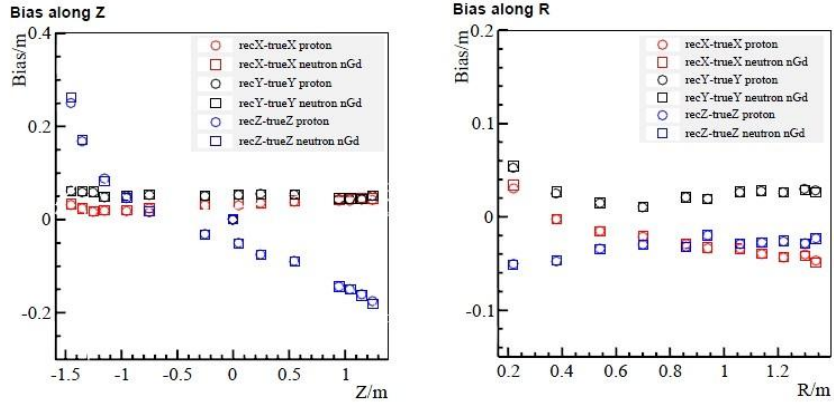

Fig.5.: Bias of the vertex reconstruction along vertical and radial direction.

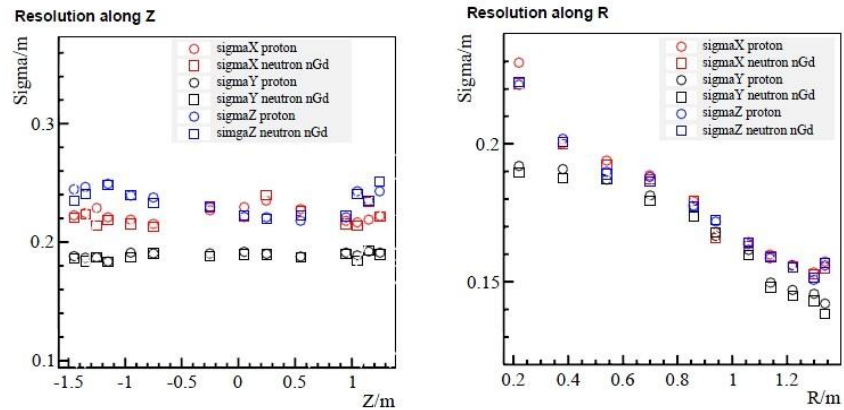

Fig.6.: Resolution of the vertex reconstruction along vertical and radial direction.

## 4. Conclusion

With the calibration data of $^{60}$Co, we optimized the key parameters of the optical model of the ML reconstruction for the Daya Bay experiment. We can predict the number of photon electron collected by each of the 192 PMTs with the optimized optical model. The uniformity of the energy reconstruction, and the bias and resolution of the vertex reconstruction of neutron events are also presented in this report.